# Dramatic increase of the onset critical temperature and critical field of elemental Sn in the form of thin nanowires

Ying Zhang[1][♠], Chi Ho Wong[2,4][♠], Junying Shen[2][♠], Sin Ting Sze[2], Yan Dong[1], Hui Xu[1], Zifeng Yan[1], Yingying Li[3], Xijun Hu[3][♣] and Rolf Lortz[2][♣]

[1] *State Key Laboratory for Heavy Oil Processing, PetroChina Key Laboratory of Catalysis, China University of Petroleum, Qingdao 266580, China*

[2] *Department of Physics, Hong Kong University of Science and Technology, Clear Water Bay, Kowloon, Hong Kong*

[3] *Department of Chemical and Biomolecular Engineering, Hong Kong University of Science and Technology, Clear Water Bay, Kowloon, Hong Kong*

[4] *Institute of Physics and Technology, Ural Federal University, Russia*

[♣] Address correspondence to kexhu@ust.hk and lortz@ust.hk .
[♠]These authors contributed equally to the article.

**Abstract**
Sn is a well-known classical superconductor on the border between type I and type II with critical temperature of 3.722K and critical field of 0.031T [1]. We show by means of specific heat and electric magneto-transport data that its critical parameters can be dramatically increased if it is brought in the form of loosely bound bundles of thin nanowires. The specific heat displays a pronounced double phase transition at 3.7K and 5.5K, which we attribute to the inner 'bulk' contribution of the nanowires and to the surface contribution, respectively. The latter is visible only because of the large volume fraction of the surface layer in relation to their bulk inner volume. The upper transition coincides with the onset of the resistive transition, while zero resistance is gradually approached below the lower transition. The large coherence length of 230nm at 0K likely actuates a Josephson coupling between adjacent neighboring nanowires and thus suppresses the effect of 1D phase fluctuations along the nanowires, and stabilizes 3D phase coherence throughout the entire network with zero resistance. A magnetic field of more than 3T is required to restore the normal state, which means that the critical field is enhanced by about two orders of magnitude with respect to Sn in its bulk form.

## I.    Introduction

Nanostructuring materials can change their physical properties dramatically. It thus provides the opportunity to tailor new materials with improved or entirely new characteristics. In the case of superconductors, the confinement of Cooper pairs in the nanometer geometry in most cases is unfavorable for properties that are suitable for applications. Nevertheless, some rare cases where the constrained geometry of nanoscale superconductors can create interesting effects have been reported. For example, a slight increase in the superconducting transition temperature was observed in nanoparticles of In [2,3], Tl [2], and Ga[4-6] and Pb nanobelts [7,8]. Furthermore, some enhancement of the critical fields [3,5,7-10] and unusual magnetoresistance oscillations have been observed [7,8]. In addition, we recently reported a huge improvement of the critical field and onset critical temperature in arrays of parallel ultrathin Pb nanowires [11]. A similar $T_c$ increase was observed more recently in Pb microspheres [12].

In reduced dimensionality, thermally induced fluctuations of the superconducting order parameter are greatly enhanced [13,14]. They lower the critical temperature dramatically below which zero resistance is found, although phase incoherent Cooper pairs may exist at much higher temperatures. In a two-dimensional superconductor, the superconducting phase transition occurs in form of a Berezinskii-Kosterlitz-Thouless transition, which occurs usually below the bulk critical temperature [15]. Quasi-one-dimensional (quasi-1D) superconductivity occurs in nanowires which are thinner than their superconducting coherence length [16,17]. In 1D superconductors, according to the Mermin-Wagner theorem, thermally induced order parameter phase slips will cause finite resistance at any $T > 0$ K [13,14]. The resistance then shows only a continuous decrease below the temperature where the Cooper pairs form [17]. Nevertheless, it has been shown theoretically [18-25] and experimentally [11,26,27] that a long-range-ordered state may be formed when many 1D superconducting nanowires are arranged in close proximity to form a regular array. Then a transverse Josephson or proximity coupling can then suppress the phase slip processes and mediates a zero resistance state at finite temperatures [28]. In this paper, we report a dramatic increase of the onset critical temperature and critical field when elemental Sn is brought into the form of networks of randomly oriented weakly coupled freestanding nanowires, while a true zero-resistance state is preserved below 2K.

## II. Sample fabrication

Sn nanowires were synthesized using surfactant as soft template by a chemical reduction process. In a typical synthesis, 0.2g sodium dodecyl sulfate (SDS) and 0.4g tin sulfate were dissolved in 80mL distilled water and heated to 303K in a water bath under stirring. After 20 minutes, a sodium borohydride solution (0.02g of $NaBH_4$ in 5 mL of distilled water) was added. The nanowires formed after 20 minutes reaction. The resulting solution was separated by centrifugation and the precipitate was washed several times with distilled water and ethanol.

The morphology and microstructures of the resulting nanowires were characterized by a field emission transmission electron microscopy (TEM, JEM-2100UHR/200kV), X-ray diffraction (XRD, Philips X'pert Pro Alpha 1 Diffractometer with Cu Kα radiation, $\lambda = 1.5406$ Å) and scanning electron microscopy (SEM, S-4800, Japan). Fig. 1a and 1b show SEM and TEM images of the Sn nanowires with 60-70 nm in diameter and about 500 nm in length. The lattice-resolved HRTEM image (Fig. 1c) clearly reveals a lattice spacing of 0.29 nm that corresponds to the (200) planes of β-Sn. The HRTEM images and the ED pattern (Fig. 1d) unambiguously demonstrate the single-crystal structure with [100] orientation.

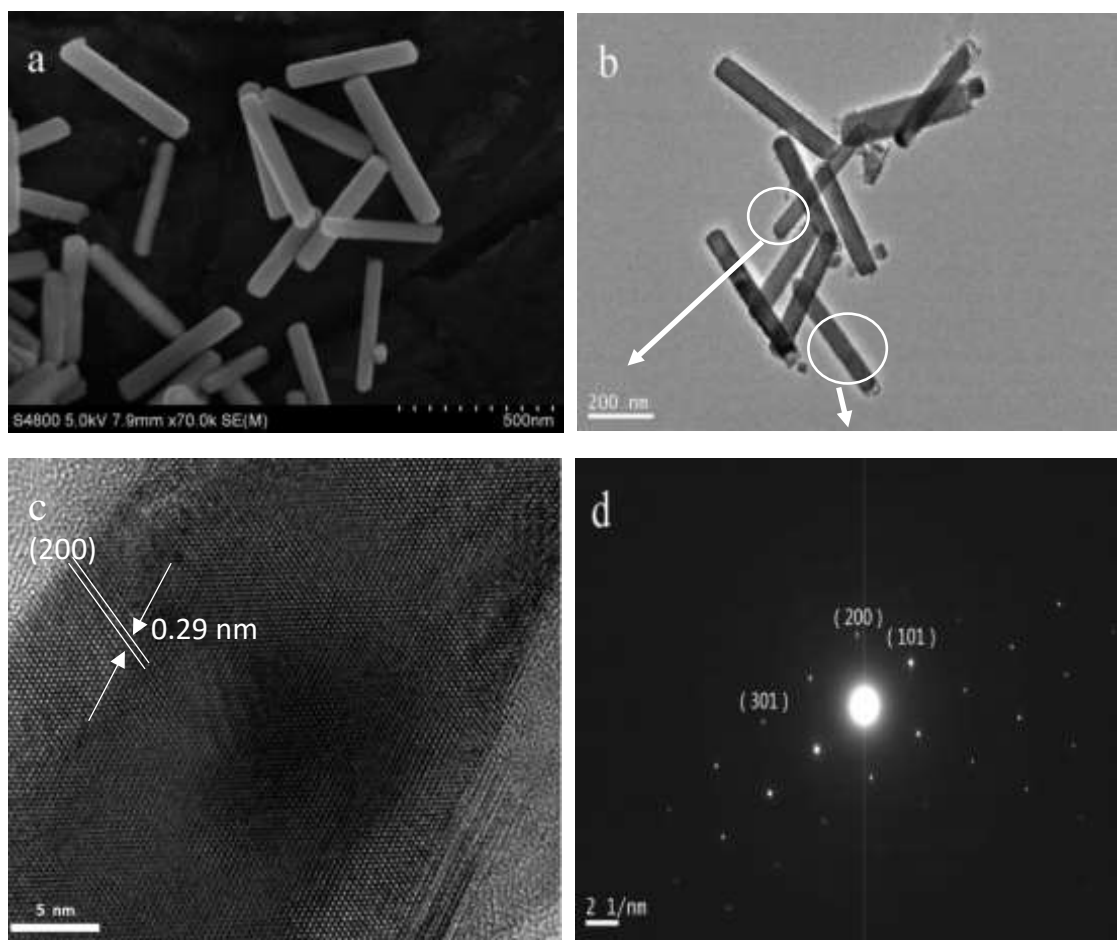

Fig. 1 (a) SEM image of Sn nanowires, (b) TEM image of Sn nanowire and its (c) HRTEM image and (d) ED pattern.

The crystal structure of Sn nanowires was recorded by powder X-ray diffraction (XRD) as shown in Fig. 2. Diffraction peaks located at 30.6 ° and 32 ° can be attributed to the (200) and (101) planes of β-Sn (JCPDS No. 01-086-2265). It shows evidence of a preferred orientation with the (200) and (101) peaks. The intensity of the (200) peak for the as-made nanowires is much higher than that of the secondary (101) peak, which confirms the result from the TEM data that the growth direction of the Sn nanowires is highly preferred along the [100] direction.

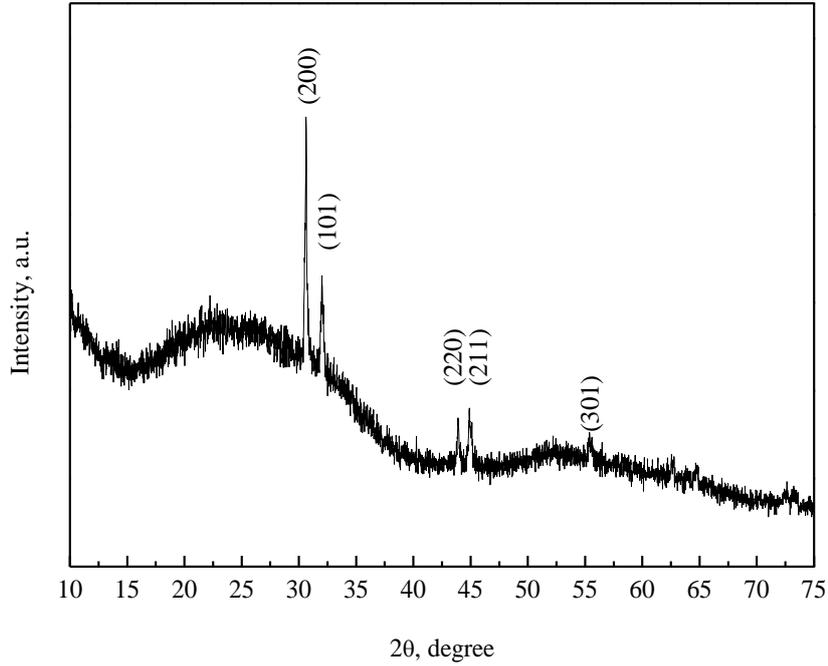

Fig. 2 Higher-angle XRD data for the Sn nanowires.

### III. Results

The specific heat of the Sn nanowires was measured with a modulated temperature AC micro-calorimetric technique [29], which provides the required high relative resolution of $\Delta C/C \approx 10^{-5}$ to resolve the tiny specific heat anomaly of these superconducting nanowires. A total mass of 100 μg sample, consistent of a few micron size grains containing the nanowire networks and mixed with insulating GE 7031 varnish as thermal compound, was mounted on the calorimeter chip. The electronic contribution was obtained in a standard way [30] by application of a magnetic field of 14T to suppress superconductivity. The tiny size of the superconducting specific heat anomaly (~1% of the total heat capacity of sample and calorimeter chip) required an extremely precise field-calibration of the thermocouple used in our calorimeter, which was done by comparing the specific heat of a silver calibration sample with and without applied magnetic field. To ensure significant precision, we only present data of the electronic specific heat derived from the difference in zero field data and 14T data. Fig. 3 shows $C^{electr}/T$ of the Sn nanowires in comparison to literature data [31] for bulk Sn.

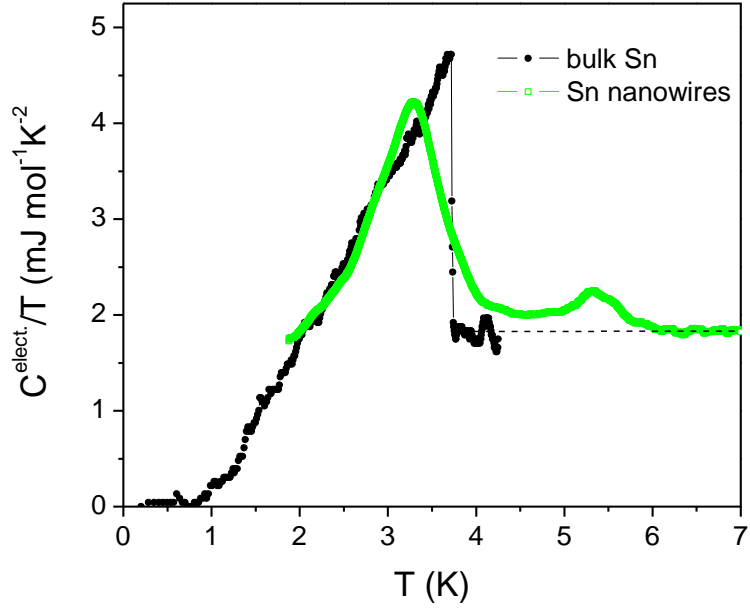

Figure 3: The electronic heat capacity $C^{electr}/T$ at the superconducting transition of Sn nanowires as a function of temperature in zero field. The black circles are literature data of bulk Sn [31].

The superconducting anomaly in the specific heat of the Sn nanowires differs significantly from that of bulk Sn and shows a distinct double transition anomaly with jumps at 3.7K and 5.5K, which both are fully suppressed by the 14T magnetic field. The center of the main transition at 3.7K coincides roughly with the much sharper jump observed in bulk Sn. However, the transition anomaly is much more continuous and shows a more symmetrical shape (albeit with a rather sharp kink at the maximum). A fluctuation tail extends well above the $T_c$ of bulk Sn up to at least 4.5K until the second smaller transition anomaly occurs with a jump at 5.5K. Below ~6K, $C_{electr}/T$ is significantly larger than the normal state Sommerfeld constant, which together with the fact that the anomaly is suppressed in parallel to the main transition by an applied field indicates its superconducting origin.

Using the two fluid model, we split the electronic heat capacity into the Sommerfeld and superconducting components and so can derive the $T_c$ distribution [32] of the Sn nanowires. The 1-$F(T)$ and d$F$/d$T$ correspond to superconducting volume fraction and the $T_c$ distribution, respectively (Fig. 4). This deconvolution of the specific heat further illustrates the large width of the superconducting transition, which extends from 1.5K to almost 6.5K, with peaks occurring in d$F$/d$T$ at 3.7K and 5.5K.

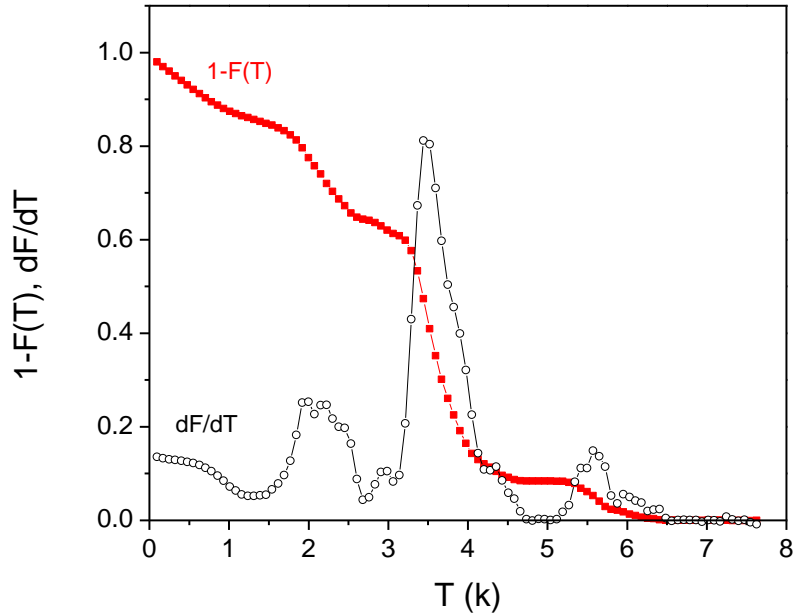

Figure 4: Superconducting volume fraction of a macroscopic sample of Sn nanowires 1-$F(T)$ and the corresponding $T_c$ distribution d$F$/d$T$ [32].

The device for resistance measurement was prepared by sputtering a layer of 5 nm Ti and a layer of 70 nm Au on a polished grain of the sample that was fixed by epoxy on an insulating substrate. Four electrodes with a distance of 3.2 μm were separated with a focused ion beam technique. The measurement was carried out by applying an AC current of 20 nA in combination with a digital lock-in amplifier.

Fig. 5 shows the electric resistance of a network of a large number of randomly aligned nanowires in the form of a small grain of a few microns in size. We present the resistance instead of resistivity, since the precise current path in this device is unknown. In view of the 1D nature of the nanowires, the resistance drops surprisingly sharp with midpoint around 4.1K. Below the main transition, the resistance remains continuously dropping until zero resistance is achieved at ~2K. Zooming to the onset of the transition (inset of Fig. 5) reveals that the resistance begins to drop significantly at 5.5K, which is 1.8K above the transition of bulk Sn. This agrees well with the results obtained from the heat capacity and confirms the superconducting origin of the second transition anomaly at 5.5K. Applying a magnetic field of 3T is required to suppress the resistive superconducting transition, which contrasts enormously to the low critical field of 0.033T of bulk Sn. Thus the resistivity data demonstrates that nanostructuring of Sn provides an enormous enhancement of both, the onset $T_c$ and the upper critical field.

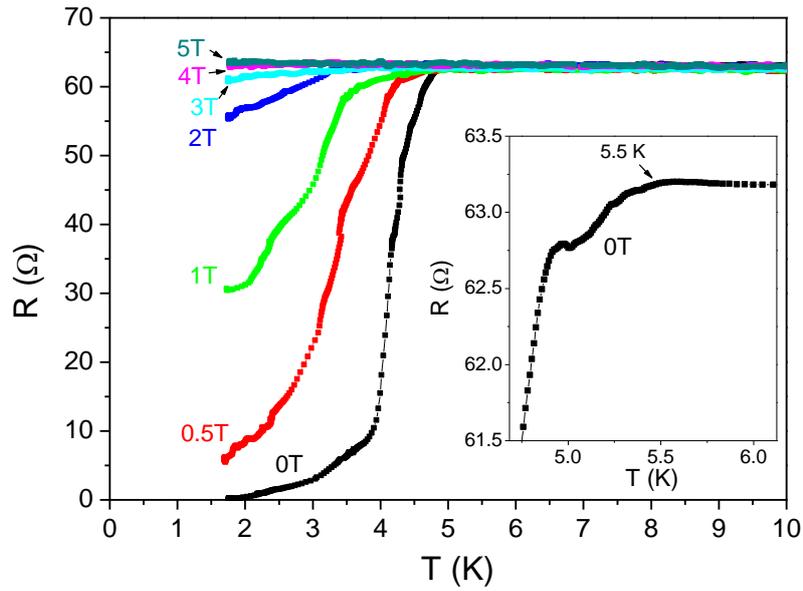

Figure 5: Electric resistance at the superconducting transition of a micrometer-size grain containing a network of many randomly aligned Sn nanowires in various magnetic fields. The inset shows an enlargement of the transition onset, which reveals that the onset of the superconducting transition is enhanced up to 5.5K.

Fig. 6 shows the magnetoresistance of Sn nanowires in field sweeps at different temperatures. At 1.4K, the magnetoresistance increases starting from the lowest fields, slowly at first and more rapidly above 0.5T, until the normal state resistance is approached at 2T. At higher temperatures, the resistance starts from finite values and the critical field is gradually lowered until a flat curve is found above 4K. This quite temperature-independent normal state resistance is likely the effect of the weak links between the nanowires.

Zooming in on the 4T data (inset of Fig. 6) shows that traces of superconductivity persist at this temperature for at least up to a broad kink at ~2.2T. This is in accordance with the weak signature of superconductivity in the zero field resistance data up to 5.5K. The fact that zero resistance is only established in the zero field limit is likely a result of the 1D nature of the nanowires, in which phase slips along the wire are causing finite resistivity at finite Meissner screening current densities [13,14].

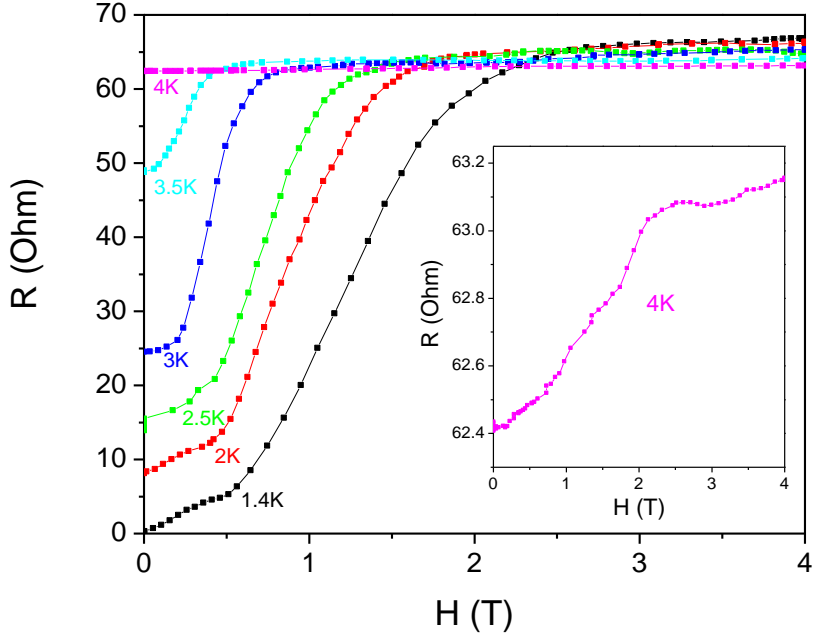

Figure 6: Magnetoresistance at various fixed temperatures.

## IV. Discussion

Our experiments show a strong enhancement of the onset $T_c$ and the upper critical field of Sn, when it is brought in the form of thin nanowires. In Fig. 7, we plot the critical field transition midpoints from Fig. 6 as well as the transition onsets (defined as the highest field where a significant deviation from the linear normal state background is observed). The magnetic field vs. temperature phase diagram obtained allows us to estimate the critical field at zero temperature. The standard Werthamer-Helfand-Hohenberg (WHH) theory [33] was used to extrapolate the data to zero temperature and estimate the upper critical field $H_{c2}$ = 1.5T. In such a 1D superconductor fluctuation effects are very strong and the onset transition is somewhat more meaningful: It represents the upper field boundary where 1D superconducting fluctuations are observed. The extrapolation shows that the critical field at which these fluctuations vanish is 3.3T. Our measurement techniques are not sensitive to probe whether the superconductor is of type-I, or type-II, but such high critical fields are usually found only in type-II superconductors, suggesting that the nanostructuring has changed the characteristics of Sn from type-I to type-II.

It is notable that the transition points of the two curves can be extrapolated to a $T_c$(0T) value of 4K, which corresponds likely to the bulk $T_c$ of the inner part of the nanowires. The additional transition of the specific heat and the onset of the resistive transition at 5.5K are thus likely not a bulk $T_c$, but rather attributed to the surface layer of the nanowires [11]. In such thin nanowires, the surface layer has a substantial volume fraction of the nanowires. The heat capacity is most sensitive to the Cooper pairing and the transition at 5.5 K represents a significant fraction of 10% of the total pairing signal (see Fig. 4). Our nanowires have an average thickness of 70 nm. This suggests that a surface layer having a thickness of up to 7 nm is influenced by the softening of the phonons near the surface. Of crucial importance here is that the nanowires are thin enough so that a significant portion of the volume is affected by the phonon softening at the surface. In addition, the curvature of the surface may represent another important ingredient for the $T_c$ enhancement in these cylindrical nanowires [34]. Without these effects, the $T_c$ of a superconductor would rather be expected to decrease with dimensionality [13,14,35].

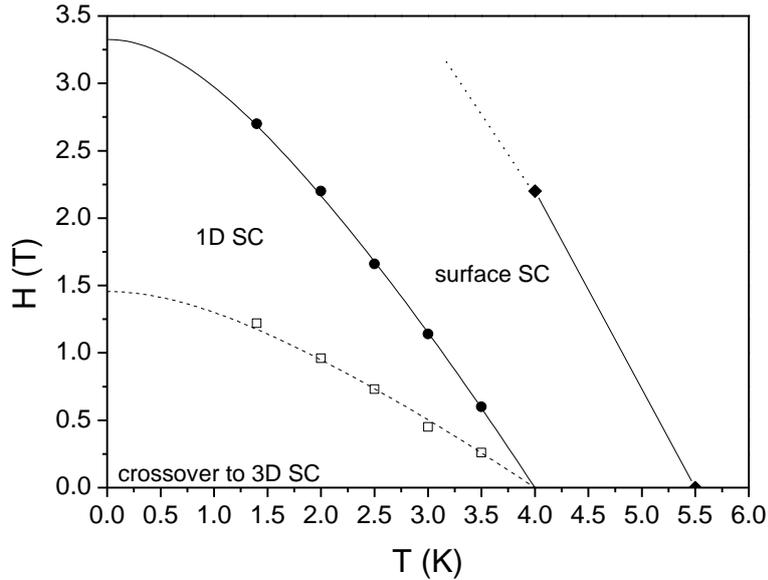

Figure 7: Magnetic field vs. temperature phase diagram showing different characteristic regions associated to surface superconductivity, 1D superconducting fluctuations and presumably a crossover to a 3D superconducting state in the entire network of nanowires within the micrometer size grains containing many bundles of loosely bound nanowires. The dashed line represents the midpoint of the broad transition in the magnetoresistance. True phase coherent 3D superconductivity is only formed below 2K in the limit of zero fields, as demonstrated by the zero resistance in the macroscopic sample. The lines are fits according to the Werthamer-Helfand-Hohenberg (WHH) theory [33].

For the enhancement of the critical field several factors must be considered. Bulk Sn is a classical superconductor at the border between type-I and type-II. A reduction of the electron mean free path in a superconductor decreases the superconducting coherence length $\xi$ and can drive it into the type-II region with larger critical field $B_{c2}$. This was shown, for example, for Pb doped with some Bi impurities [36]. The high critical field of our nanowires suggests a similar effect, caused by the confinement of the Cooper pairs in the quasi-1D structure. Moreover, a 1D material has an open Fermi surface without closed orbits. That suppresses the formation of vortices and thus the orbital limit for superconductivity [37] at which the vortex cores overlap will be greatly enhanced. Eventually, the vortex cores only penetrate the nanowire network between the wires in form of Josephson vortices [38], which have little effect on the Cooper pairs within the wires. Another factor may be the particularly strong spin orbital coupling [39] in a metallic surface state band, which has been reported to increase the upper critical field of ultrathin Pb films by two orders of magnitude [35]. That zero resistance shows that the Sn nanowires do not completely behave individually in our macroscopic powder sample, but probably show some coupling. The Josephson Effect may mediate this, if the nanowires are close to each other but without contact – or rather point contacts when they are in physical contact. The resistivity is thus likely most sensitive to this phase ordering effect, which establishes global phase coherence throughout the network of nanowires (probed by our experiment on the micrometer scale). Global phase coherence throughout the network is required for the resistance to drop to zero, while phase incoherent 1D fluctuations that apparently form below 5.5K, only cause the slight decrease in resistivity without a strong magnetoresistance effect in the high temperature regime. This coupling obviously then establishes a bulk phase-coherent state in the macroscopic sample, which may occur in a similar way as observed in Pb nanowires in SBA-15 matrices [11] or in intrinsically quasi-1D superconductors as $Tl_2Mo_6Se_6$ [40] or $Sc_3CoC_4$ [41].

We can thus identify 3 different regions in the phase diagram: an area in which surface superconductivity is observed in a thickness of about 7 nm below ~5.5K, a region of quasi-1D superconductivity between 2K and 4K and a crossover to a bulk 3D phase-coherent superconducting state with zero resistance. Note that the latter is only established entirely in the limit of zero fields, while phase slips in the order parameter cause a finite resistance even at very low magnetic field strengths. This shows that the coupling of the nanowires is rather weak in the network. It may probably be weakened by their random orientation, while a larger field range of zero resistance was found in arrays of ultrathin parallel-aligned Pb nanowires [11].

**Conclusion**

We demonstrated that the superconducting characteristics of networks of freestanding Sn nanowires are substantially modified by nanostructuring, with a 1.8K increase of the onset $T_c$ and an increase in the upper critical field by an enormous factor of 100. The surface and the curvature mediated phonon softening in a 7 nm thin surface layer is suggested to be the reason to push the onset $T_c$ from 3.72K to 5.5K, while the giant enhancement of the upper critical field likely originates from the absence of closed orbits around the quasi-1D Fermi surface and strong spin-orbit coupling effects. A similar effect was observed previously in the in ultrathin Pb nanowires [11], making our findings suggest that such an enhancement in the superconducting properties may be a fairly common effect of classical superconducting materials induced by nanostructuring.

**Acknowledgements**

This work was financially supported by the National Natural Science Foundation of China (No. 51271215), the Innovative Exploratory Grant of Hong Kong University of

Science and Technology (No. IEG14EG02PG) and grants from the Research Grants Council of the Hong Kong Special Administrative Region, China (No. FSGRF14SC25, No. FSGRF15SC07).